\documentclass[aps,twocolumn,prl,superscriptaddress]{revtex4-1}

\usepackage{amsmath}
\usepackage{amssymb}
\usepackage{color}
\usepackage{hyperref}

\usepackage{graphicx}

\DeclareMathAlphabet{\mathitbf}{OML}{cmm}{b}{it}

\newcommand{\dbar}{{\,\mathchar'26\mkern-12mu d}}

\newcommand{\sFrac}[2]{{\textstyle\frac{#1}{#2}}}

\begin{document}
%\title{Effect of preparation protocol on the density of vibrational modes in model glasses}

\title{Effect of instantaneous and continuous quenches on the density of vibrational modes in model glasses}
% \author{Edan Lerner}
% \affiliation{Institute for Theoretical Physics, University of Amsterdam, Science Park 904, 1098 XH Amsterdam, The Netherlands}
% \author{Eran Bouchbinder}
% \affiliation{Chemical Physics Department, Weizmann Institute of Science, Rehovot 7610001, Israel}

\author{Edan Lerner}
\affiliation{Institute for Theoretical Physics, University of Amsterdam, Science Park 904, 1098 XH Amsterdam, The Netherlands}
\author{Eran Bouchbinder}
\affiliation{Chemical Physics Department, Weizmann Institute of Science, Rehovot 7610001, Israel}

\begin{abstract}
Computational studies of supercooled liquids often focus on various analyses of their ``underlying inherent states" --- the glassy configurations at zero temperature obtained by an infinitely-fast (instantaneous) quench from equilibrium supercooled states. Similar protocols are also regularly employed in investigations of the unjamming transition at which the rigidity of decompressed soft-sphere packings is lost. Here we investigate the statistics and localization properties of low-frequency vibrational modes of glassy configurations obtained by such instantaneous quenches. We show that the density of vibrational modes grows as $\omega^\beta$ with $\beta$ depending on the parent temperature $T_{0}$ from which the glassy configurations were instantaneously quenched. For quenches from high temperature liquid states we find $\beta\!\approx\!3$, whereas $\beta$ appears to approach the previously-observed value $\beta\!=\!4$ as $T_0$ approaches the glass transition temperature. We discuss the consistency of our findings with the theoretical framework of the Soft Potential Model, and contrast them with similar measurements performed on configurations obtained by continuous quenches at finite cooling rates. Our results suggest that any physical quench at rates sufficiently slower than the inverse vibrational timescale --- including all physically-realistic quenching rates of molecular or atomistic glasses --- would result in a glass whose density of vibrational modes is universally characterized by $\beta\!=\!4$.

\end{abstract}

\maketitle

%\section{introduction}

\emph{Introduction.--} Instantaneous quenches of high-temperature configurations into their so-called ``underlying inherent states" are a prevalent practice in computational studies of disordered materials \cite{Sastry2000,Heuer2008,karmakar_lengthscale,widmer2008irreversible,exist,ohern2003}. One conspicuous example of this methodology is found in the large body of numerical work dedicated to the unjamming scenario (see e.g.~\cite{liu_review} and references therein), in which simple models of soft repulsive spheres are regularly employed. In these studies, packings of soft spheres at zero temperature are conventionally generated by instantaneous quenches from some random, high-energy disordered states, and later subjected to~various structural analyses \cite{matthieu_PRE_2005,van_hecke,everyone,mizuno2017continuum} and/or perturbations \cite{breakdown, xu2017instabilities}. % Oftentimes, the results of these structural analyses are proposed to represent glassy solids created by quenches of liquids at physical rates. 

A similar methodology is also extensively utilized in computational investigations of the glass transition \cite{Sastry2000,Heuer2008,karmakar_lengthscale,widmer2008irreversible}, whose structural origin remains a highly-debated topic in condensed-matter physics \cite{Debenedetti2001,Cavagna200951}. Instantaneous quenches that map an equilibrium configuration to a zero-temperature glassy state were first put forward by Stillinger and Weber \cite{Stillinger1983}, and subsequently utilized by many others \cite{Sastry2000,Heuer2008,karmakar_lengthscale,widmer2008irreversible}, with the general assumption that the structural properties of the inherent states are indicative in some quantitative way of the dynamics of the supercooled configurations from which they are mapped. 

A clear advantage of analyzing the structure of glassy inherent states over equilibrium configurations is the ability to cleanly and quickly extract structural observables while avoiding the difficulties that stem from thermal-fluctuations-induced noise, and from the broad spectrum of relaxation times that characterizes these systems. Instantaneous quenches are considered to be unrealistic idealizations of the physical cooling process by which glasses are formed. However, it is regularly assumed that \emph{generic} properties of the resulting glasses remain unaffected by such protocols. This uncontrolled assumption overlooks the potential physical artifacts involved in performing instantaneous quenches in computational studies of the structural properties of glassy materials.

%Some of the many theoretical frameworks for the glass transition such as random first-order theory \cite{rfot}, frustration-limited domain theory \cite{Tarjus2005}, or locally favored structures \cite{Coslovich2007,PatrickRoyall2008} postulate the emergence of some sort of order (sometimes referred to as `amorphous order' \cite{Biroli2008}) which is responsible for the immense slowing down of relaxational dynamics as supercooling proceeds towards the glass transition temperature $T_g$.  Computer simulations play a key role in providing supporting evidence for these theoretical approaches. In particular, some attempts to link relaxational dynamics to structure in supercooled liquids opt for the analysis of the inherent states that underlie equilibrium supercooled configurations \cite{widmer2008irreversible,karmakar_lengthscale}. 

In this Rapid Communication we question the common practice of investigating glassy states that were instantaneously-quenched from high-temperature configurations, and subsequently deducing conclusions about generic glasses formed via physically-realistic protocols. We focus in particular on the statistical and structural properties of low-frequency vibrational modes measured in ensembles of inherent states created by an instantaneous quench of configurations equilibrated at various parent temperatures $T_0$. Recent studies of several structural glass forming models \cite{modes_prl,SciPost2016,mizuno2017continuum} (see additional comments about the relation between \cite{modes_prl} and the present work in \cite{modes_prl_erratum}) identified a population of quasilocalized low-frequency glassy vibrational modes whose density $D(\omega)$ grows from vanishing frequencies $\omega\!\to\!0$ as $D(\omega)\!\sim\!\omega^4$. These modes are either measured below the lowest Goldstone modes' frequency \cite{modes_prl,SciPost2016}, or identified by classifying vibrational modes according to their degree of localization \cite{mizuno2017continuum}. Similar findings for a three-dimensional Heisenberg spin glass in a random field were put forward in \cite{parisi_spin_glass}. Here we show that the \emph{functional form} of the density of low-frequency vibrational modes can be affected by instantaneous quenches, and, under some conditions, displays deviations from the $\omega^4$ law.  %This work constitutes the first steps in systematically assessing the robustness of the $\omega^4$ law tvariations in the way glasses are quenched from a melt. 
%Our results call for care to be taken in attempts to draw conclusions about generic glasses from studies of the structural properties of glasses formed by instantaneous quenches. 
We contrast our findings with measurements performed in ensembles of inherent states created by a continuous quench at a cooling rate $\dot{T}$ through our model systems' glass transition temperature. We find that the $\omega^4$ law is robust to very rapid but not overdamped quenches, suggesting that inertia plays a key role in the self-organizational processes that occur as systems tumble down the multi-dimensional potential energy landscape during their quench into a glassy solid.

\emph{Models and methods.---} Here we briefly review the models and methods used in this work; a detailed description of our model, methods and preparation protocols can be found in the Supplemental Material (SM) \cite{SM}. We employ a binary mixture of point-like particles in three dimensions (3D) that interact via a purely repulsive inverse-power-law potential. In what follows physical observables (temperatures, frequencies, lengths, etc.) are understood as expressed in terms of the relevant microscopic units as defined in the SM. For visualization purposes alone we rescale frequency axes by a scale $\omega_0$, see figure captions. We chose to simulate systems of $N\!=\!2000$ particles for which the linear size of the box is slightly larger than the localization length of low-frequency glassy modes (estimated in our model at about 10 particle diameters \cite{modes_prl}), but still small enough such that Goldstone modes are sufficiently suppressed, allowing ample exposure of vibrational modes that occur below the lowest Goldstone mode frequency~\cite{modes_prl}.

Ensembles of inherent states were created by collecting a large number of independent equilibrium configurations from each parent temperature $T_0$, and evolving each one of these forward in time under fully overdamped dynamics $\dot{\vec{x}} \!\propto\! -\frac{\partial U}{\partial \vec{x}}$ until convergence, where $\vec{x}$ denotes particles' coordinates and $U$ the potential energy. We have also created ensembles of continuously-quenched glasses, starting from independent equilibrium configurations at $T\!=\!1.00$, followed by a quench at a prescribed quench rate $\dot{T}$. Each of the constructed ensembles consists of 10,000 glassy samples, which ensures statistical convergence, see SM for further details. 
%A zero temperature glassy sample was deemed converged once the ratio of the mean magnitude of the net force on each particle to the pressure dropped below $10^{-10}$.

 %for these quenches and for our equilibrium runs we employed a Berendsen thermostat \cite{berendsen}; we fixed the Berendsen thermostat time-parameter in our equilibrium simulations at $\tau_{\mbox{\tiny Ber}}\!=\!10.0$, which is more than an order of magnitude larger than the vibrational time scale of our model. For continuous quenches at high rates we are forced to decrease $\tau_{\mbox{\tiny Ber}}$, otherwise the heat in the system is unable to catch up with the continuously-decreasing target temperature. We chose $\tau_{\mbox{\tiny Ber}}$ to be the largest possible (i.e.~we aimed at minimizing its effect), while maintianing the ability to quench at the desired rate \cite{footnote}.

\begin{figure}[!ht]
\centering
\includegraphics[width = 0.495\textwidth]{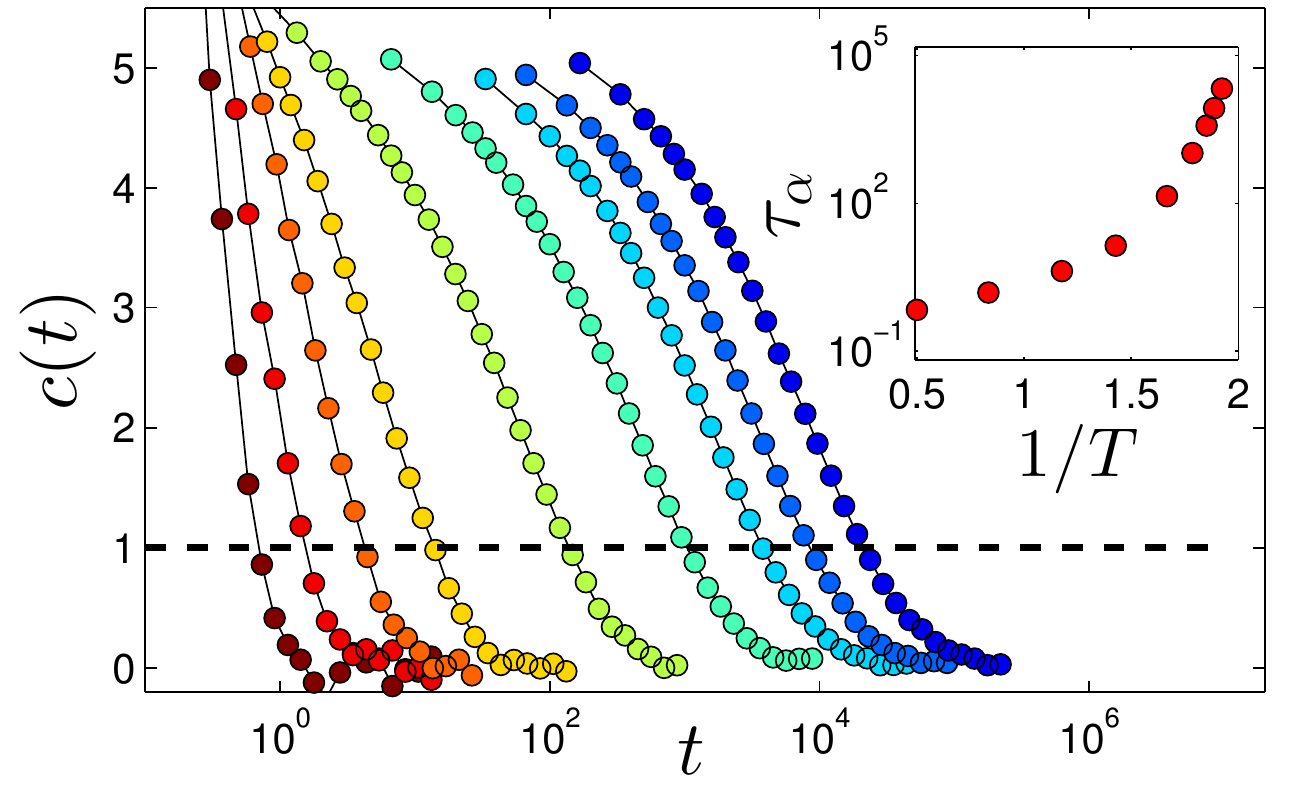}
\caption{\footnotesize (color online) Stress autocorrelation function (see text for definition) measured in equilibrium simulation runs at temperatures $T\!=\!2.00, 1.20, 0.85, 0.70, 0.60, 0.56, 0.54, 0.53,$ and $0.52$, decreasing from left to right. Inset: the relaxation times $\tau_\alpha$ vs.~$1/T$, determined by $c(\tau_\alpha)\!=\!1$, as indicated by the dashed horizontal line of the main panel.}
\label{fig1}
\end{figure}

\emph{Results.---} Our model system exhibits the conventional phenomenology of computer glass forming models. In Fig.~\ref{fig1} we demonstrate the slowing down in the relaxational dynamics upon supercooling of our model by monitoring the stress autocorrelation function $c(t)\!\equiv\! N\langle\sigma(t)\sigma(0)\rangle$ measured at various equilibrium runs at temperatures $T$. Here $\sigma\!\equiv\!\frac{1}{V}\frac{\partial U}{\partial\gamma}$, $V$ is the volume of the simulation cell, $U$ is the potential energy and $\gamma$ is a simple shear strain. The angular brackets denote an average over the time-translationally-invariant signals of the stress from our equilibrium simulations. The inset of Fig.~\ref{fig1} shows the relaxation time $\tau_\alpha$ vs.~$1/T$; relaxation times are estimated via $c(\tau_\alpha)\!=\!1$, as indicated by the horizontal dashed line. The computer glass transition temperature of our model is estimated at $T_g\!\approx\!0.5$, where the relaxation time $\tau_\alpha(T_g)\!\approx\!10^5$.

%%%%%%%%%%%%%%%%%%%%%%%%%%%%%%%%%%%%%%%%%%%%%%%%%%%%%%%
%%%%%%%%%%%%%%%%%%%%%%%%%%%%%%%%%%%%%%%%%%%%%%%%%%%%%%

%%%%%%%%%%%%%%%%%%%%%%%%%%%%%%%%%%%%%%%%%%%%%%%%%%%%%%%
\begin{figure}[!ht]
\centering
\includegraphics[width = 0.48\textwidth]{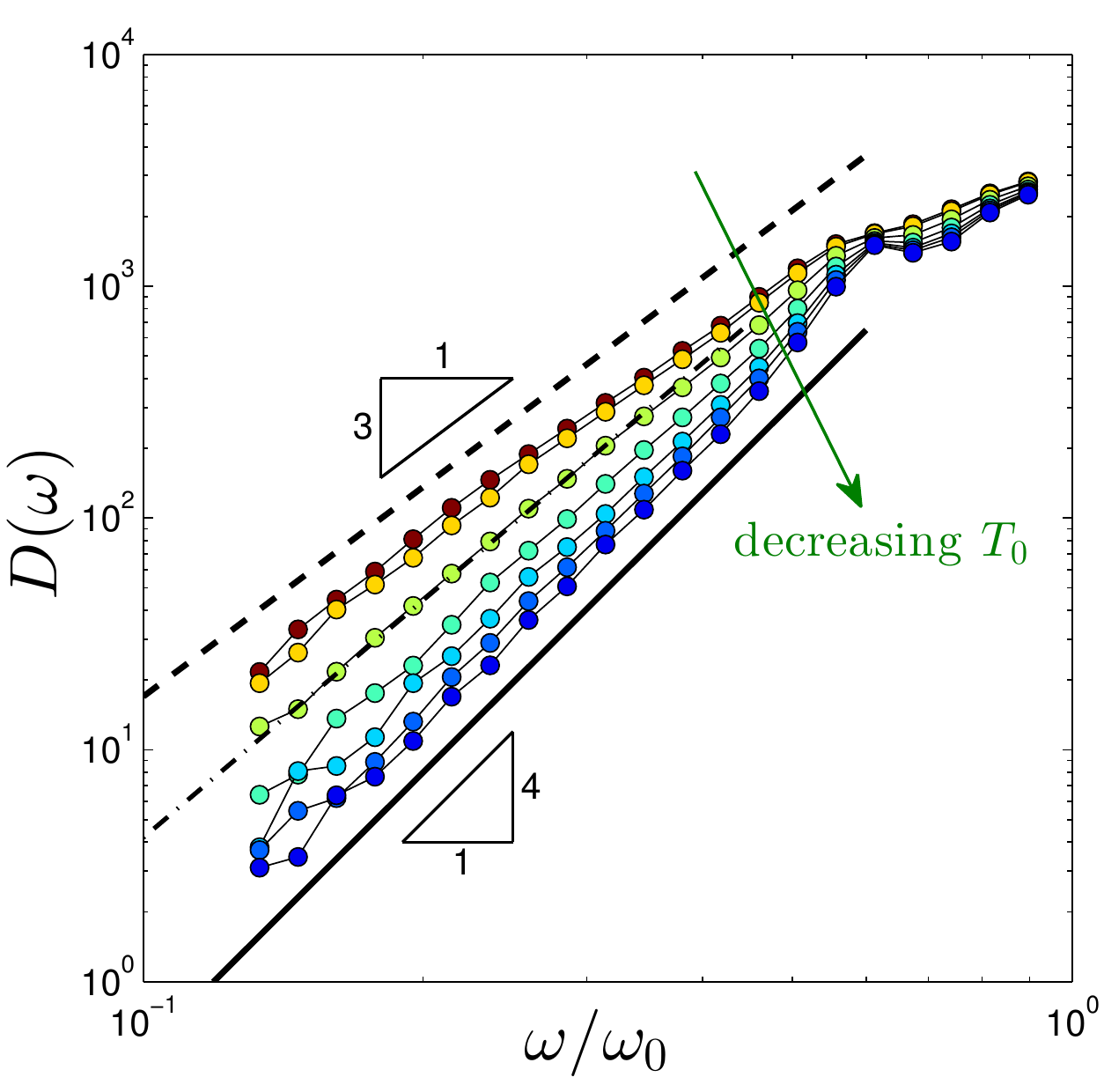}
\caption{\footnotesize (color online) Density of vibrational modes $D(\omega)$ measured in the ensembles of glassy samples quenched from the parent temperatures $T_0\!=\!2.00, 0.70, 0.60, 0.56, 0.54, 0.53,$ and $0.52$, decreasing from top to bottom. The frequency axis is scaled by $\omega_0\!=\!3.0$ for visualization purposes. The dash-dotted line fitted to the $T_0\!=\!0.60$ data set corresponds to $D(\omega)\!\sim\!\omega^{3.4}$.}
\label{fig2}
\end{figure}
%%%%%%%%%%%%%%%%%%%%%%%%%%%%%%%%%%%%%%%%%%%%%%%%%%%%%%

We next turn to the investigation of the statistics and properties of vibrational modes in the different ensembles of instantaneously quenched glasses. Each such ensemble was obtained by an instantaneous quench of independent configurations that were equilibrated at some parent temperature $T_0$. In Fig.~\ref{fig2} we show the low-frequency tails of the density of vibrational modes $D(\omega)$ measured in all the ensembles of glassy samples that were instantaneously quenched from parent temperatures as indicated by the figure caption. We find that $D(\omega)\!\sim\!\omega^\beta$ with $3<\beta\!\le\!4$ for all ensembles, and $\beta\!\to\!4$ as $T_0\!\to\! T_g$. These data demonstrate that it is not only that the high $T_0$ inherent states possess more soft glassy vibrational modes, but that the actual functional form of the vibrational modes' distribution function depends explicitly on $T_0$, at least up to the vicinity of the accessible equilibrium temperatures using conventional simulation methods. We emphasize at this point that our goal is not to accurately estimate the precise numerical value of the scaling exponents that characterize the density of vibrational modes. Our aim is rather to identify trends in the observed exponents upon systematically varying the preparation protocol of the glassy samples. 

%%%%%%%%%%%%%%%%%%%%%%%%%%%%%%%%%%%%%%%%%%%%%%%%%%%%%%%
\begin{figure}[!ht]
\centering
\includegraphics[width = 0.495\textwidth]{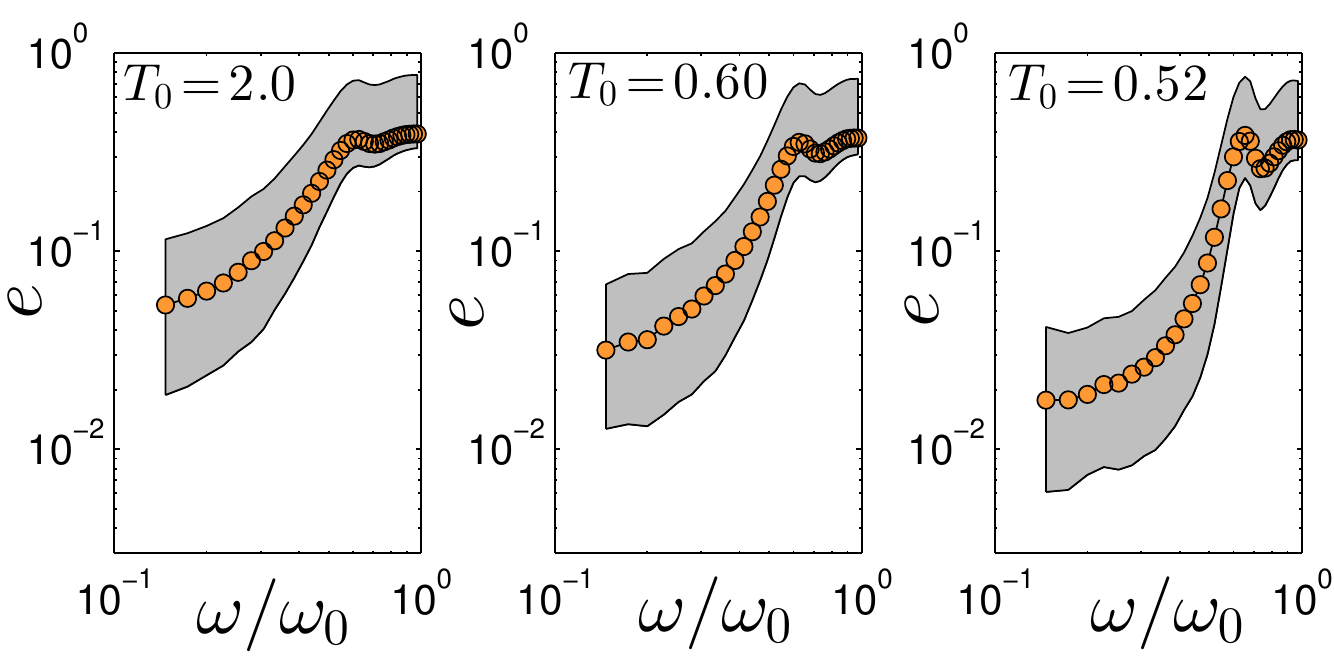}
\caption{\footnotesize (color online) Participation ratio $e$ of vibrational modes vs.~frequency $\omega$, for glassy samples instantaneously quenched from a parent temperature $T_0$ as indicated in the figure. The shaded gray areas cover the 2nd-9th deciles of data, and the circles represent the mean participation ratio binned over frequency. The frequency axes are scaled by $\omega_0\!=\!3.0$ for visualization purposes. Stronger localization is observed as $T_0\!\to\!T_g$. }
\label{fig3}
\end{figure}
%%%%%%%%%%%%%%%%%%%%%%%%%%%%%%%%%%%%%%%%%%%%%%%%%%%%%%

In Fig.~\ref{fig3} we show the means and the 2nd-9th deciles of the participation ratio $e$ binned over frequency $\omega$. The participation ratio of a vibrational mode $\hat{\Psi}$, defined as $e\!\equiv\!\big( N\sum_i (\hat{\Psi}_i\!\cdot\!\hat{\Psi}_i)^2 \big)^{-1}$, is a simple measure of the degree of localization of a mode: the more localized a mode is, the smaller its participation ratio is expected to be. In \cite{modes_prl, SciPost2016} it has been shown that the participation ratio of low-frequency glassy modes scales as $N^{-1}$, indicating that they are quasilocalized \cite{footnote2}. Fig.~\ref{fig3} here shows that the degree of localization of low-frequency glassy modes increases for deeper supercooling, consistently with the findings of \cite{modes_prl} that show a decrease in the participation ratio of low-frequency glassy modes for slower cooling rates. The data indicate that the transition of the mean participation ratio from the Goldstone modes' value to the low-frequency plateau (shown clearly for a much larger data set in \cite{SciPost2016}) is faster in ensembles created by an instantaneous quench from deeply supercooled solids. Furthermore, the rapid crossover in the localization properties of modes with increasing frequencies suggests that the exponent $\beta$ can only be read off $D(\omega)$ below frequencies that are roughly a third of the lowest Goldstone mode frequency, see e.g.~the data in Fig.~\ref{fig4} below. We note that the crossover from quasilocalized, glassy modes at low frequencies to the first Goldstone modes is broader for smaller $N$; this can be seen, for instance, in Fig.~1 of Ref.~\cite{SciPost2016}. In the SM we show that increasing the system size does not, however, appear to have a substantial effect on our results, which reinforces the statement that the crossover broadening in our systems of $N\!=\!2000$ does not effect our conclusions.

It is natural to contrast our results for instantaneously quenched glasses with similar measurements in glassy samples formed by a continuous quench into solids at finite quench rates. In Fig.~\ref{fig4} we show the density of vibrational modes of systems quenched at rates $\dot{T}$ as described in the legend. Each such quench was preformed from initial equilibrium configurations at temperature $T\!=\!1.00$ (see SM for details). We find $\beta\!=\!4$ at rates $\dot{T}\!<\!10^{-2}$, as shown by the continuous lines. For higher rates, $\beta$ appears to decrease, and for an infinitely-fast quench from $T\!=\!1.00$ we find $\beta\!\approx\!3.3$ as indicated by the dashed line.

%%%%%%%%%%%%%%%%%%%%%%%%%%%%%%%%%%%%%%%%%%%%%%%%%%%%%%%
\begin{figure}[!ht]
\centering
\includegraphics[width = 0.495\textwidth]{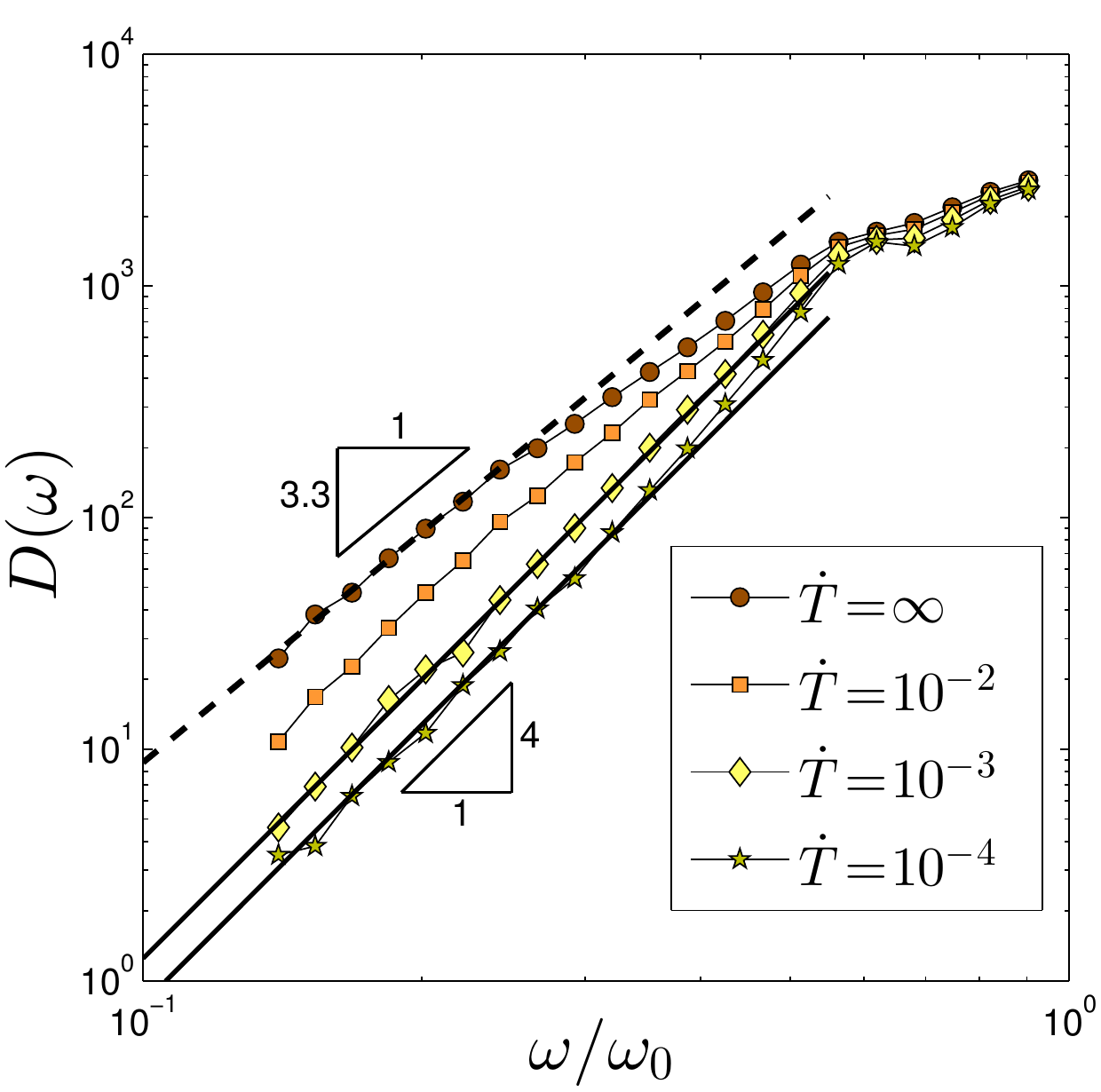}
\caption{\footnotesize (color online) Density of vibrational modes $D(\omega)$ measured in emsembles of glassy samples quenched continuously at rates as indicated by the legend, starting from equilibrium configurations at $T\!=\!1.00$. Both continuous lines correspond to the $\omega^4$ law. The frequency axis is scaled by $\omega_0\!=\!3.0$ for visualization purposes. }
\label{fig4}
\end{figure}
%%%%%%%%%%%%%%%%%%%%%%%%%%%%%%%%%%%%%%%%%%%%%%%%%%%%%%

\begin{figure}[!ht]
\centering
\includegraphics[width = 0.495\textwidth]{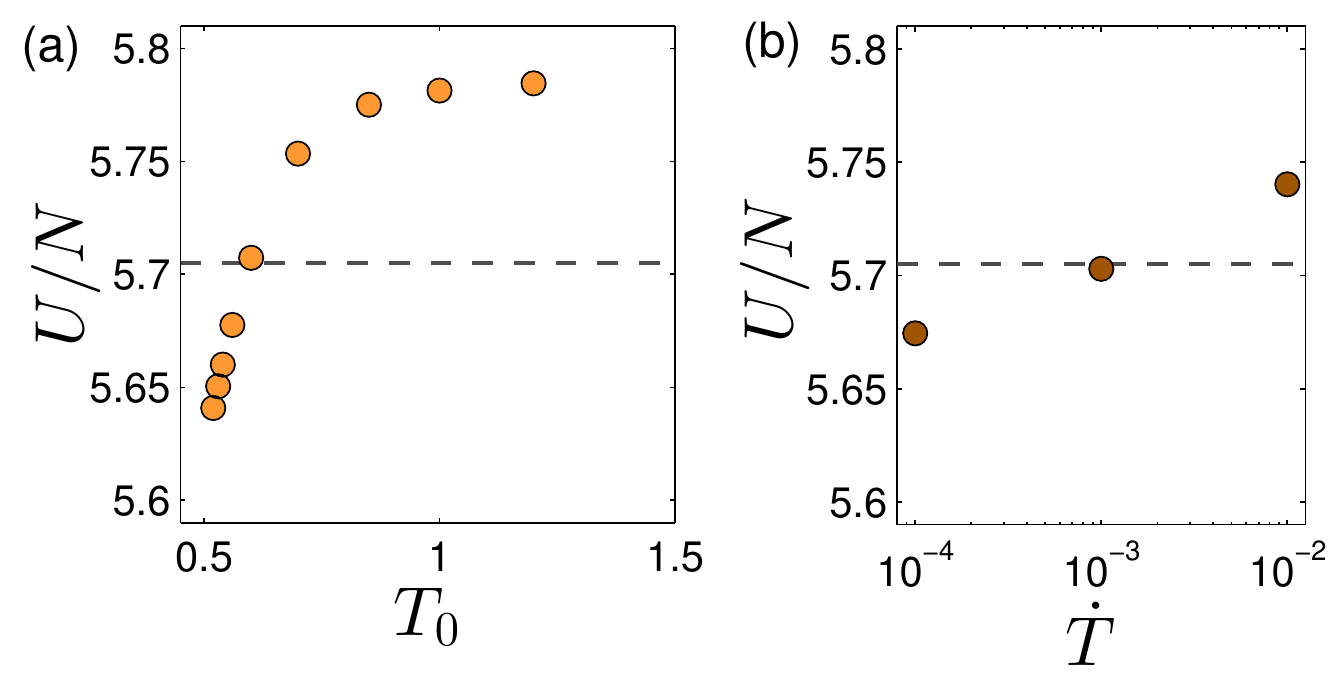}
\caption{\footnotesize (color online) Potential energy per particle of glassy samples averaged over (a) ensembles created by instantaneous quenches from the parent temperature $T_0$, and (b) ensembles created by continuous quenches at quench rates $\dot{T}$. The dashed horizontal line shows that the $T_0\!=\!0.60$ ensemble and the $\dot{T}\!=\!10^{-3}$ ensemble have very similar energies per particle, see text for further discussion.}
\label{iso_energy_fig}
\end{figure}

In Fig.~\ref{iso_energy_fig} we plot the average potential energy per particle of the ensembles of instantaneously-quenched and continuously-quenched glassy samples. Interestingly, we find that the mean energy per particle of glasses quenched at the highest continuous rate for which $\beta\!=\!4$ is observed ($\dot{T}\!=\!10^{-3}$, see Fig.~\ref{fig4}) is the same as for instantaneously quenched samples from the parent temperature $T_0\!=\!0.60$, up to less than a percent. However, in the latter ensemble we clearly find $\beta\!<\!4$, see dash-dotted line in Fig.~\ref{fig2}. This observation of two ensembles with the same inherent state energies but different $\beta$ indicates that inertia that is present during the continuous quenches, but absent in the instantaneous quenches, plays an important role in the self-organizational processes that determine the fine details of the microstructure of the resulting glasses.

%%%%%%%%%%%%%%%%%%%%%%%%%%%%%%%%%%%%%%%%%%%%%%%%%%%%%%%
%%%%%%%%%%%%%%%%%%%%%%%%%%%%%%%%%%%%%%%%%%%%%%%%%%%%%%

In order to explore the implications of our results for realistic glasses, we cast our reported observables into dimensionless numbers. We start with forming an atomistic timescale by considering the shear-wave speed $c_s\!\equiv\!\sqrt{\mu/\rho}\!\approx\!4$, with an athermal shear modulus \cite{lutsko} $\mu\!\approx\!15$ and a mass density $\rho\!=\!0.82$, and dividing it by an atomistic length $a_0\!\approx\!1.0$ to find $c_s/a_0\!\approx\!4$. We next take $T_g\!\approx\!0.5$ as a characteristic temperature scale, such that a dimensionless quench rate is formed as $\frac{\dot{T}a_0}{T_gc_s}$. Our findings suggest that for dimensionless quench rates lower than a crossover value $10^{-3}$, or alternatively, for quench rates $\dot{T}\!\lesssim\!10^{-3}T_gc_s/a_0$, the density of vibrational modes exhibits the $\omega^4$ law. To compare to physical glasses, e.g.~metallic glasses, we take $T_g\!\approx\!500$K, $c_s\!\approx\!10^3\mbox{m/sec}$, and $a_0\!\approx\!10^{-9}\mbox{m}$ \cite{Wang200445}, from which we conclude that glasses quenched at rates $\dot{T}\!\lesssim\!10^{11}\mbox{K/sec}$ would exhibit the $\omega^4$ law. The fastest rates that these materials can be quenched are typically on the order of $10^7\mbox{K/sec}$ (for quasi-2D ribbons; for bulk glasses the fastest cooling rates are slower), some 4 decades slower than our estimated crossover rate. This comparison essentially implies that any laboratory glass formed by quenching a melt would follow the $\omega^4$ law.

\emph{Summary and discussion.--} In this Rapid Communication we have shown that the low-frequency tails of the density of vibrational modes of computer glasses created by an instantaneous quench have qualitatively different features compared to glasses created by a continuous quench. Our results suggest that the presence of inertia is important for the structural relaxation that occurs during quenches which leads to more stable glassy structures with less low-frequency vibrational modes. This suggestion is consistent with the results of Salerno \emph{et al}.~\cite{salerno_robbins}, who showed that upon reducing the inertia in the microscopic dynamics of sheared model glasses, the nature of avalanches of plastic activity, which depends in turn on the abundance of soft glassy modes, can change dramatically. Similar findings were reported in \cite{Barrat_inertia}. If indeed the presence of inertia in the microscopic dynamics is key in determining the low-frequency spectra, it would be of interest to observe whether overdamped glasses such as emulsions or foams, or computer glasses generated in simulations that employ Brownian dynamics, exhibit observable qualitative differences in their spectra compared to their inertial counterparts.

Our results call for caution when attempts are made to establish general conclusions about glassy solids from studies of model glasses that are created by instantaneous quenches from high temperature liquid states. For instance, it is common practice in studies of the unjamming point to create packings of soft spheres by instantaneous quenches. While the qualitative features of the scaling of most mechanical observables with respect to the distance to the unjamming point do not seem to depend on the protocol with which packings are generated, our results suggests that the density of vibrational modes of those packings might not be representative of the spectra of glasses created by physical quenches. 

It is interesting to attempt to relate our findings to the predictions of the Soft Potential Model \cite{soft_potential_model,soft_potential_model_01,soft_potential_model_02,Chalker2003}. This theoretical framework assumes that a glass can be decomposed into small subsystems, each possessing a quasilocalized soft glassy mode. Focusing on such a typical subsystem, and assuming that particles are displaced a distance $s$ along the soft mode associated with that subsystem, this framework suggests that if the energy in the vicinity of $s\!=\!0$ satisfies $U(s)\!\ge\! U(0)$, then $D(\omega)$ is expected to grow as $\omega^4$. However, relaxing this constraint results in a different prediction, namely that $D(\omega)\!\sim\!\omega^3$ \cite{Chalker2003}. The condition that the energy only grows in the vicinity of $s\!=\!0$ can be viewed as a stability condition; in instantaneously quenched glasses the overdamped nature of the quench makes it possible to form barely-stable glasses that would possess local soft potentials $U(s)$ that have deeper minima at $s\!\ne\!0$ compared to $U(0)$, e.g.~asymmetric double well potentials \cite{soft_potential_model_02}. Creating such unstable structures in slowly quenched glasses is much less likely. According to the discussed framework, one may hypothesize that $\beta\!=\!3$ should be observed in glasses created by an instantaneous quench; we indeed find $\beta$ very close to 3 in samples that were instantaneously quenched from very high temperatures, see Fig.~\ref{fig2}.

In this work we followed the simple approach of \cite{modes_prl} and investigated the density of vibrational modes in small, three dimensional model glasses, in which Goldstone modes are sufficiently suppressed to expose a population of quasilocalized soft glassy vibrational modes. This approach is, however, still limited in terms of the range of soft glassy modes' frequencies that can be probed, due to hybridizations with extended Goldstone modes at higher frequencies, as can be seen in Fig.~\ref{fig3}. It is therefore of interest to investigate these issues using frameworks that overcome the issue of hybridization with Goldstone modes e.g.~\cite{SciPost2016,thermal_energies,pinning_brito,parisi_spin_glass,manningDisentangling}, allowing one to probe the density of quasilocalized excitations up to higher frequencies.

A key question to be addressed in future research is whether extremely slow quench rates can result in glasses with $\beta\!>\!4$. Recent developments \cite{berthier_swap_hard_spheres, berthier_prx} in the computational research of structural glasses allow one to equilibrate a particular model glass well below what is possible using conventional molecular dynamics or Monte Carlo methods. The new methodology introduced in \cite{berthier_swap_hard_spheres, berthier_prx} will be certainly useful in addressing this question.

\textit{Acknowledgments.--} We thank Gustavo D\"uring, Eric DeGiuli, and Matthieu Wyart for fruitful discussions.  % E.B.~acknowledges support from the Israel Science Foundation (Grant No.~712/12), the Harold Perlman Family Foundation and the William Z. and Eda Bess Novick Young Scientist Fund. 

\vskip 0.17cm

%\bibliography{references_lerner}

\begin{thebibliography}{37}
\bibitem [{foo()}]{footnote2}%
  \BibitemOpen
\bibfield  {journal} {  }\href@noop {} {\bibinfo  {journal} {The term
  quasilocalized is used to describe low-frequency glassy modes since their
  amplitude has been shown \cite{modes_prl} to decay as a power-law away from
  their core}\ }\BibitemShut {NoStop}%
\end{thebibliography}
\begin{thebibliography}{4}
\bibitem{footnote1}
We note that the low-frequency glassy modes are not truly localized; in \cite{modes_prl1} they were shown to be quasilocalized, namely their amplitude decays in the far field as $r^{-2}$.
\end{thebibliography}

\newpage
\begin{center}
        \textbf{\large Supplemental Material for:\\``Effect of instantaneous and continuous quenches on the density of vibrational modes in model glasses"}
\end{center}

%%%%%%%%%%%%%%%%%%%%%%%%%%%%%%%%%%%%%%%%%%%%%%%%%%%%%%%%%%%%%%%%%%%%%%%%%%%%%%%%%
%%%%%%%%%%%%%%%%%%%%%% these lines of code handle the concatenation %%%%%%%%%%%%%
%%%%%%%%%%%%%%%%%%%%%%%%%%%%%%%%%%%%%%%%%%%%%%%%%%%%%%%%%%%%%%%%%%%%%%%%%%%%%%%%%
 
\setcounter{equation}{0}
\setcounter{figure}{0}
\setcounter{section}{0}
\setcounter{table}{0}
\setcounter{page}{1}
\makeatletter
\renewcommand{\theequation}{S\arabic{equation}}
\renewcommand{\thefigure}{S\arabic{figure}}
\renewcommand*{\thepage}{S\arabic{page}}
\renewcommand{\bibnumfmt}[1]{[S#1]}
\renewcommand{\citenumfont}[1]{S#1}

In this supplemental material (SM) we describe the models and numerical methods employed in our work. In addition, we show data concerning the statistical convergence of our results, and the absence of finite-size effects in our spectra calculations.

% \subsubsection*{Plastic modes and displacement reponses to local dipolar forces}
% In our work we show examples of plastic modes (see Fig.~4 in the main text) calculated in harmonic disc packings (see details below). In Fig.~\ref{compare_fig} of this SM we show that the spatial structure of these plastic modes resembles the displacement response to dipolar forces, calculated as described in \cite{breakdown}. This resemblance suggests that the core-size of plastic modes is controlled by the same lengthscale below which continuum linear elasticity appears to breakdown, as shown in \cite{breakdown}.  

%%%%%%%%%%%%%%%%%%%%%%%%%%%%%%%%%%%%%%%%%%%%%%%%%%%%%%%
% \begin{figure}[!ht]
% \centering
% \includegraphics[scale = 0.57]{compare_point_response_plastic_modes.pdf}
% \caption{\footnotesize (color online). Panels {\bf a)}-{\bf c)} display the same plastic modes shown in Fig.~4 of the main text, that pertain to harmonic packings of $N=90000$ discs at pressures $p=10^{-2}$, $p=10^{-4}$, and $p=10^{-6}$, respectively. Panels {\bf d)}-{\bf f)} display displacement responses to local dipolar forces (see \cite{breakdown} for details about this calculation), in the same packings of panels {\bf a)}-{\bf c)}.}
% \label{compare_fig}
% \end{figure}
%%%%%%%%%%%%%%%%%%%%%%%%%%%%%%%%%%%%%%%%%%%%%%%%%%%%%%

\section{Models and numerical methods}

\subsection{Model definitions}

We employed a 50:50 binary mixture of `large' and `small' particles of equal mass $m$ enclosed in a cubic three dimensional box of linear size $L$, interacting via a radially-symmetric purely repulsive inverse power-law pairwise potential of the form
\begin{equation}
\varphi(r_{ij}) = \left\{ \begin{array}{ccc}\varepsilon\left[ \left( \sFrac{\lambda_{ij}}{r_{ij}} \right)^n + \sum\limits_{\ell=0}^q c_{2\ell}\left(\sFrac{r_{ij}}{\lambda_{ij}}\right)^{2\ell}\right]&,&\sFrac{r_{ij}}{\lambda_{ij}}\le x_c\\0&,&\sFrac{r_{ij}}{\lambda_{ij}}> x_c\end{array} \right.,
\end{equation}
where $r_{ij}$ is the distance between the $i^{\mbox{\tiny th}}$ and $j^{\mbox{\tiny th}}$ particles, $\varepsilon$ is a microscopic energy scale, and $x_c$ is the dimensionless distance for which $\varphi$ vanishes continuously up to $q$ derivatives. Distances are measured in terms of the interaction lengthscale $\lambda$ between two `small' particles, and the rest are chosen to be $\lambda_{ij}\!=\!1.18\lambda$ for one `small' and one `large' particle, and $\lambda_{ij}\!=\!1.4\lambda$ for two `large' particles. The coefficients $c_{2\ell}$ are given by
\begin{equation}
c_{2\ell} = \frac{(-1)^{\ell+1}}{(2q-2\ell)!!(2\ell)!!}\frac{(n+2q)!!}{(n-2)!!(n+2\ell)}x_c^{-(n+2\ell)}\,.
\end{equation}
We chose the parameters $x_c\!=\!1.48, n\!=\!10$, and $q\!=\!3$. The density was set to be $N/L^3\!=\!0.82\lambda^{-3}$, with the total number of particles $N\!=\!2000$ used in the majority of the numerical simulations performed (see additional discussion about system size effects below). Time is expressed in terms of $\tau_0\!\equiv\!\sqrt{m\lambda^2}/\varepsilon$, temperature in terms of $\varepsilon/k_B$ with $k_B$ the Boltzmann constant, quench rates in terms of $\varepsilon/(k_B\tau_0)$, stresses in terms of $\varepsilon/\lambda^3$, and vibrational frequencies in terms of~$\tau_0^{-1}$. In plots of the density of vibrational modes here and in the main text, we rescaled the $x$-axis by an arbitrary scale $\omega_0$ as reported in the figure caption, for visualization purposes. 

\subsection{Thermostating}

Temperature was controlled using a Berendsen thermostat scheme \cite{berendsen1}, which amounts to multiplying the momentum vector of each particle at every integration step by a factor $C_{\mbox{\tiny Ber}}$ calculated as
\begin{equation}
C_{\mbox{\tiny Ber}} \equiv \sqrt{1+\frac{\delta t}{\tau_{\mbox{\tiny Ber}}}\frac{T - \tilde{T}(t)}{\tilde{T}(t)}}\,,
\end{equation}
where $\tilde{T}(t)\!\equiv\!\frac{m}{3N}\sum_iv_i^2(t)$ is the instantaneous temperature with $v_i$ the magnitude of the velocity of the $i^{\mbox{\tiny th}}$ particle, and $\delta t$ is the numerical integration step, chosen to be $0.005\tau_0$ for $T\!\le\!1.0\varepsilon/k_B$ and $0.001\tau_0$ for $T\!>\!1.0\varepsilon/k_B$. This thermostating scheme requires chosing a time parameter $\tau_{\mbox{\tiny Ber}}$ which controls the rate at which heat is injected into or removed from the system; chosing a large value of $\tau_{\mbox{\tiny Ber}}$ is preferable in order to minimize the intervention of the thermostat with the purely Newtonian dynamics of the simulation. In particular, the dynamics reduces to Newtonian in the limit $\tau_{\mbox{\tiny Ber}}\!\to\!\infty$. In our equilibrium runs we chose $\tau_{\mbox{\tiny Ber}}\!=\!10.0\tau_0$, and made sure that systems were equilibrated during several $\tau_{\mbox{\tiny Ber}}$ before collecting statistics. This is only relevant for high temperature runs, as for lower temperatures the $\alpha$-relaxation time $\tau_\alpha$ (see main text for definition and measurements) becomes much larger than $\tau_{\mbox{\tiny Ber}}$. 

%%%%%%%%%%%%%%%%%%%%%%%%%%%%%%%%%%%%%%%%%%%%%%%%%%%%%%%
\begin{figure}[!ht]
\centering
\includegraphics[width=1.0\linewidth]{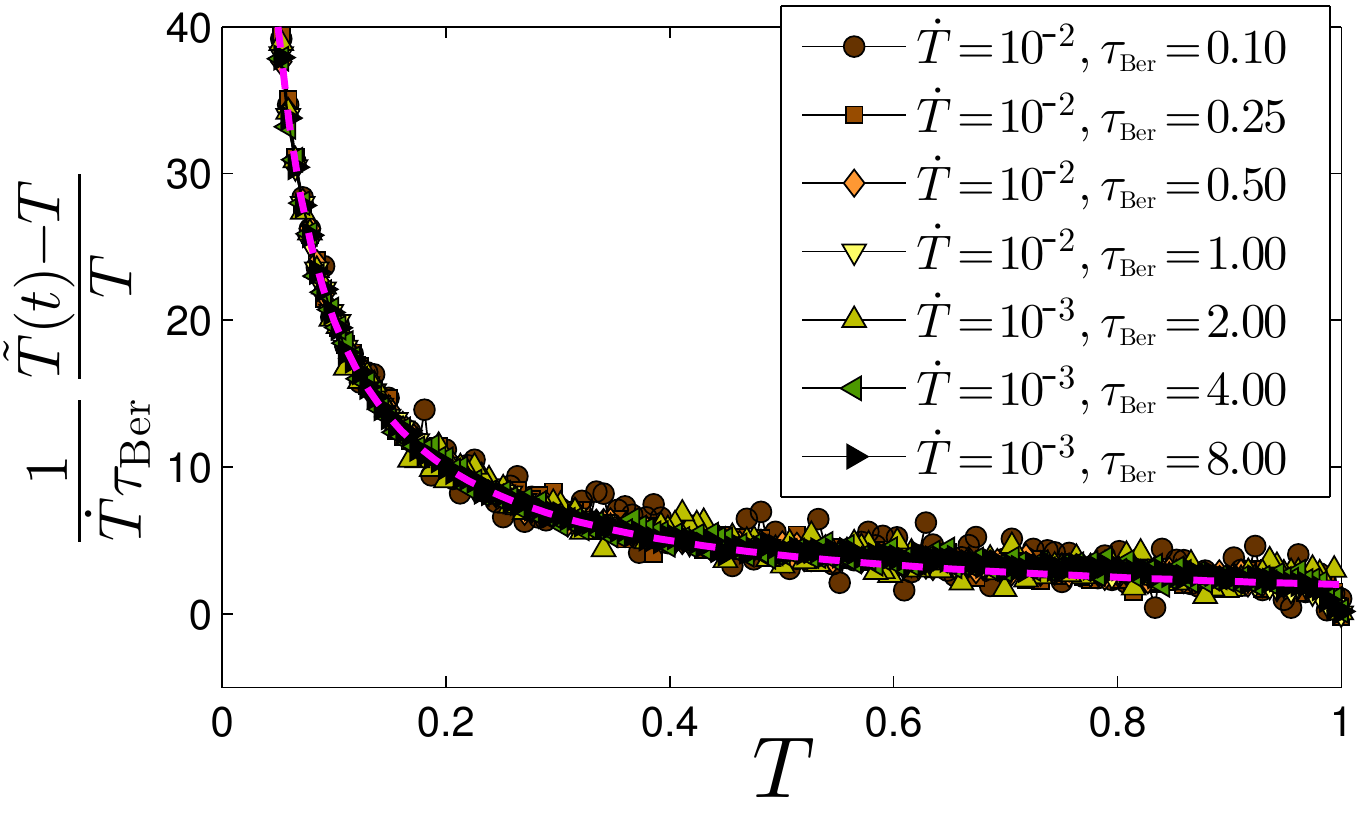}
\caption{\footnotesize The relative deviations of the instantaneous temperature $\tilde{T}(t)$ from the target temperature $T(t)$, rescaled by the inverse of the product of the quench rate $\dot{T}$ and the Berendsen thermostat time parameter $\tau_{\mbox{\tiny Ber}}$. The dashed magenta line represents the function $2/T$.  }
\label{relative_deviations_fig}
\end{figure}
%%%%%%%%%%%%%%%%%%%%%%%%%%%%%%%%%%%%%%%%%%%%%%%%%%%%%%

\subsection{Instantaneous and finite-rate quenches}

We employed two schemes for generating glassy samples. In the first scheme, we equilibrated statistically independent systems at various parent temperatures $T_0$, and then evolved time using fully overdamped dynamics. During the overdamped dynamics we calculate a characteristic interaction force scale $\bar{f} \equiv \sqrt{ \sum_\alpha f_\alpha^2/N}$ and a characteristic net force scale $\bar{F} \equiv \sqrt{ \sum_i |\vec{F}_i|^2/N}$, where $\alpha$ labels a pair of interacting particles, $f_\alpha \equiv -\frac{\partial \varphi}{\partial r_\alpha}$ is the force exerted between the $\alpha^{\mbox{\tiny th}}$ pair, $\vec{F}_i \equiv -\frac{\partial U}{\partial\vec{x}_i}$ is the net force experienced by the $i^{\mbox{\tiny th}}$ particle. The system was deemed a glass at mechanical equilibrium once the ratio $\bar{F}/\bar{f}$ dropped below $10^{-10}$. The integration step used for these overdamped runs was $\delta t/10$, where $\delta t$ was chosen as reported above. 

For the second scheme we started from high temperature equilibrium liquid states at $T\!=\!1.0\varepsilon/k_B$, which is roughly $2T_g$ for our model. We then used the Berendsen thermostat to cool down our high temperature liquid states at a perscribed rate $\dot{T}$, until the final temperature of $0.05\varepsilon/k_B\!\approx\! T_g/10$ is reached. Finally, overdamped dynamics were employed as described above to remove the remaining heat. % from the system. 

%%%%%%%%%%%%%%%%%%%%%%%%%%%%%%%%%%%%%%%%%%%%%%%%%%%%%%%
\begin{figure}[!ht]
\centering
\includegraphics[width=1.0\linewidth]{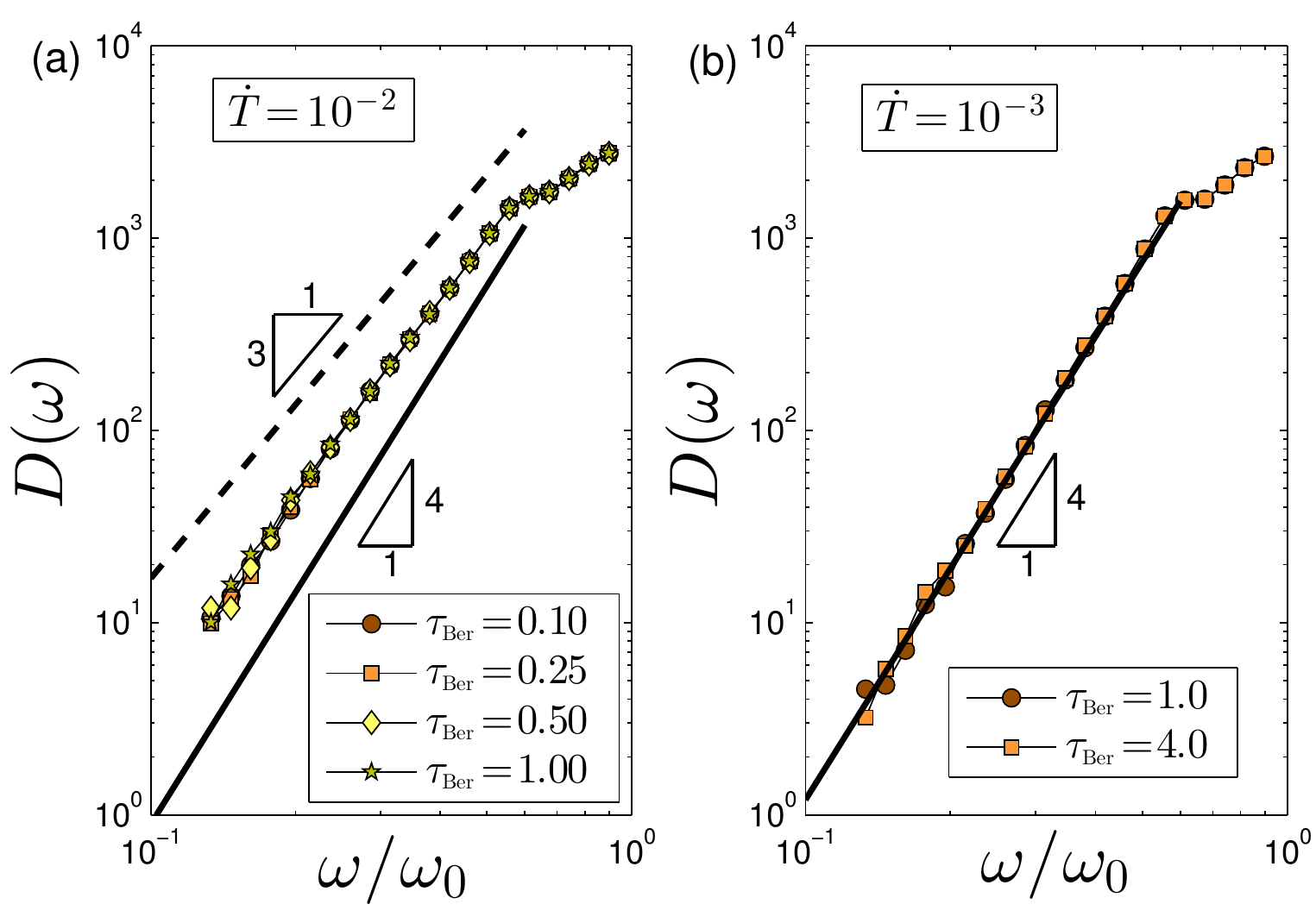}
\caption{\footnotesize Low-frequency tails of the density of vibrational modes $D(\omega)$ measured in ensembles of continuously quenched samples at quench rates $\dot{T}\!=\! 10^{-2}$ (left panel) and $\dot{T}\!=\! 10^{-3}$ (right panel), see text for further discussion. The $x$-axis of both panels is rescaled by the frequency scale $\omega_0\!=\!3.0$ for visualization purposes. }
\label{tau_ber_fig}
\end{figure}
%%%%%%%%%%%%%%%%%%%%%%%%%%%%%%%%%%%%%%%%%%%%%%%%%%%%%%

\subsubsection*{Chosing the time parameter $\tau_{\mbox{\tiny Ber}}$}

During the continuous quenches as described above the instantaneous temperature $\tilde{T}(t)$ lags behind (or strictly speaking, above) the target temperature $T(t)$, which, in this protocol, is also a function of time. We empirically observe that the lag closely follows
\begin{equation}
\frac{\tilde{T}(t)-T(t)}{T(t)} \approx \frac{2\tau_{\mbox{\tiny Ber}}\dot{T}}{T(t)}\,.
\end{equation}
as shown in Fig.~\ref{relative_deviations_fig} above. This means that given a quench rate $\dot{T}$, the time parameter $\tau_{\mbox{\tiny Ber}}$ must be chosen to be small enough for the instantanous temperature of the system to closely follow the target temperature. Following this constraint, we chose the time parameter $\tau_{\mbox{\tiny Ber}}$ for the continuous quenches such that the relative deviation of the instantaneous temperature from the target temperature at $T(t)\!=\! T_g$ remains smaller than 3\%. This translates to $\tau_{\mbox{\tiny Ber}}\!=\!0.5\tau_0$ for $\dot{T}\!=\!10^{-2}\varepsilon/(k_B\tau_0)$, and $\tau_{\mbox{\tiny Ber}}\!=\!4.0\tau_0$ for $\dot{T}\!=\!10^{-3}\varepsilon/(k_B\tau_0)$. For the slower cooling rate of $\dot{T}\!=\!10^{-4}$ we chose $\tau_{\mbox{\tiny Ber}}\!=\!10.0\tau_0$ which is itself sufficiently larger than the microscopic timescale $\tau_0$ to have any observable effects.

To check for effects of the time parameter $\tau_{\mbox{\tiny Ber}}$ on our results, we have carried out independent continous quench runs using different values of $\tau_{\mbox{\tiny Ber}}$. In Fig.~\ref{tau_ber_fig} we show data for the low-frequency tails of the density of vibrational modes (see details below) obtained for different time parameters $\tau_{\mbox{\tiny Ber}}$ and different quench rates $\dot{T}$. We find that our results are largely insensitive to variations of the time parameter $\tau_{\mbox{\tiny Ber}}$; following the notation of the main text, i.e.~$D(\omega)\!\sim\!\omega^\beta$, we find $\beta\!=\!4$ for $\dot{T}\!=\! 10^{-3}$, and $\beta$ slightly smaller than 4 for $\dot{T}\!=\! 10^{-2}$. We clearly cannot rule out that $\beta\!\to\!4$ as $\omega\!\to\!0$. However, these data demonstrate that our conclusions are not biased by our choice of the time parameter $\tau_{\mbox{\tiny Ber}}$.

%%%%%%%%%%%%%%%%%%%%%%%%%%%%%%%%%%%%%%%%%%%%%%%%%%%%%%%
\begin{figure}[!ht]
\centering
\includegraphics[width=1.0\linewidth]{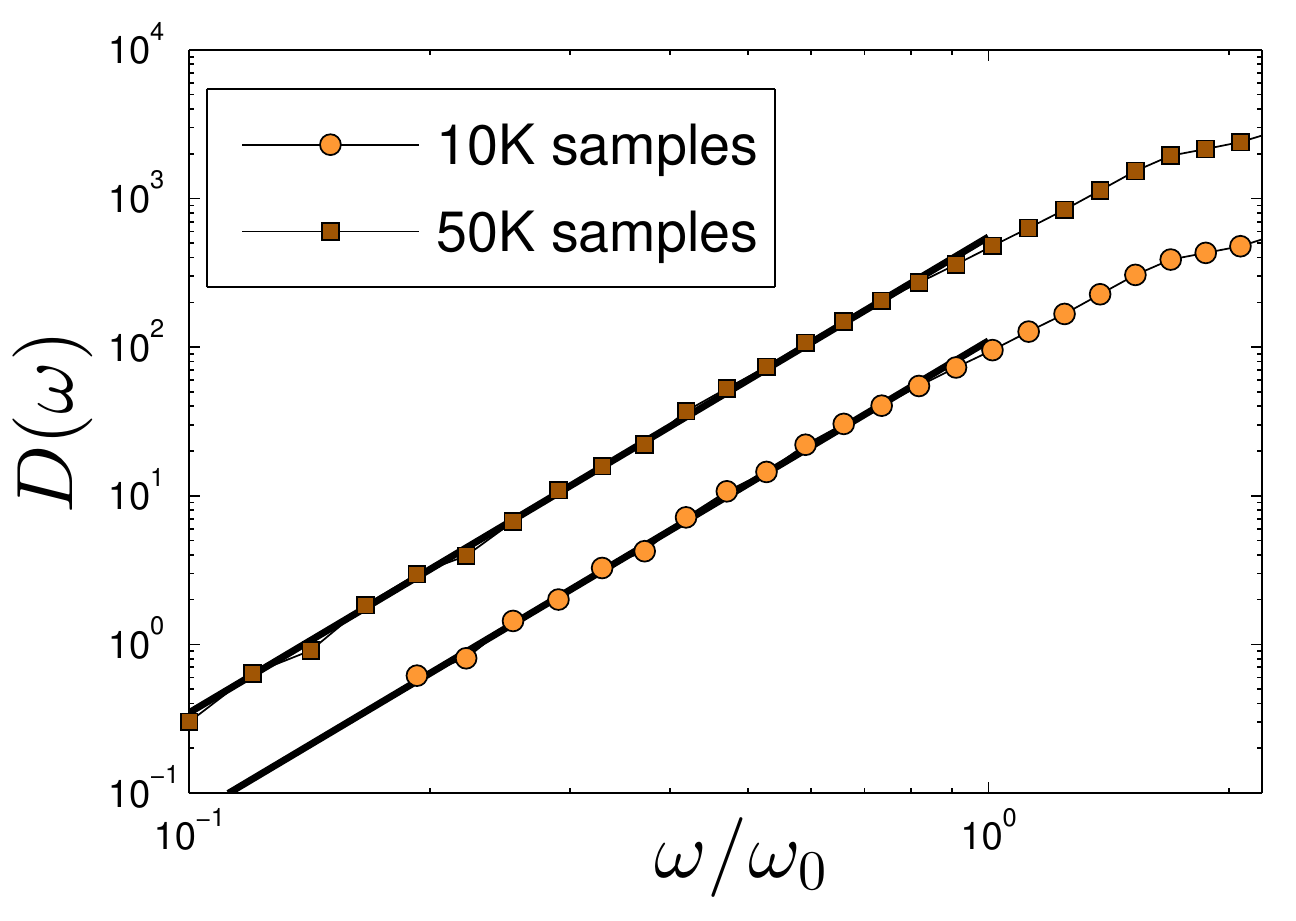}
\caption{\footnotesize Low-frequency tails of the density of vibrational modes $D(\omega)$ measured in two ensembles of 10,000 (circles) and 50,000 (squares) instantaneously quenched samples from the parent temperature $T_0\!=\!2.00$, shifted vertically for visibility. The larger data set reveals modes with lower frequencies. We see no signs of a crossover in the low-frequency tails. Here $\omega_0\!=\!1.0$}
\label{statistically_converged}
\end{figure}
%%%%%%%%%%%%%%%%%%%%%%%%%%%%%%%%%%%%%%%%%%%%%%%%%%%%%%

\subsection{\small Spectra calculations}

Normal mode analyses were carried out using the numerical analysis software MATLAB \cite{matlab1}. We calculated the first (lowest) 100, 200, and 400 modes for each glassy sample of systems of size $N\!=\!2000, 4000$ and $10,000$, respectively. 

%%%%%%%%%%%%%%%%%%%%%%%%%%%%%%%%%%%%%%%%%%%%%%%%%%%%%%%
\begin{figure*}[!ht]
\centering
\includegraphics[width=0.9\linewidth]{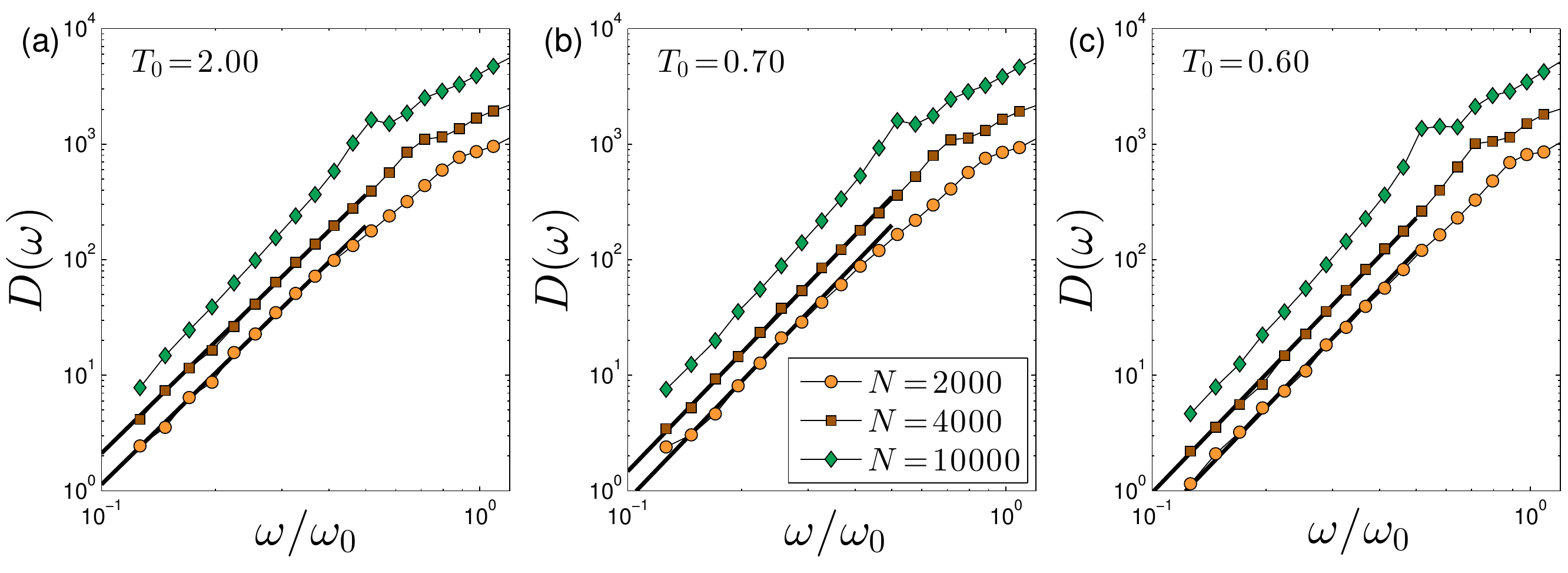}
\caption{\footnotesize Low-frequency tails of the density of vibrational modes $D(\omega)$ for calculated for various system sizes, measured in ensembles of glassy solids quenched instantaneously from equilibrium configurations at the parent temperatures of (a) $T_0\!=\! 2.00$, (b) $T_0\!=\! 0.70$ and (c) $T_0\!=\! 0.60$, see text for further discussion. For visualization purposes, the $x$-axis of all panels is rescaled by the frequency scale $\omega_0\!=\!1.9$, and the distributions are shifted vertically. The pair of continuous thick lines desribe the scaling $\omega^{3.2}$ in panel (a), and $\omega^{3.4}$ in panels (b) and (c).}
\label{finite_size_fig}
\end{figure*}
%%%%%%%%%%%%%%%%%%%%%%%%%%%%%%%%%%%%%%%%%%%%%%%%%%%%%%

\section{Statistical convergence}

In this section we provide evidence indicating that our data sets are sufficiently large and therefore statistically converged. We test in particular the instantaneous quench ensemble with parent temperature $T_0\!=\!2.00$, and independently generate an additional data set which is 5 times larger, i.e. we instantaneously quenched 50,000 independent equilibrium configurations, using the methods described above. The results for the density of vibrational modes are displayed in Fig.~\ref{statistically_converged}. While the larger ensemble reveals lower frequency modes, we do not see any signs of a crossover to a different scaling at lower frequencies, and conclude therefore that our usual ensembles of 10,000 glassy samples are sufficiently large to allow for reasonable statistical convergence.

\section{Finite size effects}

In this section we present and discuss data for the density of vibrational modes calculated in a variety of system sizes, in order to assess to what degree our observations are effected by the finite-sizes of our simulations. We focus on the instantaneous quench protocol, as reported in Fig.~2 of the main text. Before presenting and discussing our data, we remind the reader that in \cite{modes_prl1} it was shown that the possibility to observe the non-Debye low-frequency tail of the density of vibrational modes relies on delicately tuning the system sizes studied; on one hand, the spatial structure of soft glassy modes in generic models of structural glasses ---such as the one studied here--- are characterized by a localization length~\cite{footnote1} which is of the order of 10 particle sizes, as shown in \cite{modes_prl1}. The system sizes considered must be large enough to accommodate this localization length. On the other hand, the number of vibrational modes with frequencies smaller than the lowest frequency phonon vanishes with increasing the system size. This statement can be made more quantitative; to this aim, we assume that the density of vibrational modes grows as $D(\omega)\!\sim\!\omega^\beta$. For a given preparation protocol, the number of glassy vibrational modes $n_g$ that appear below the lowest frequency phonon in a single sample of linear size $L$ follows
\begin{equation}
n_g \sim L^\dbar\int_0^{L^{-1}}D(\omega)d\omega \sim L^{\dbar - \beta-1}\,,
\end{equation}
where $\dbar$ denotes the spatial dimension. Since $\beta\!\ge\!3$ is observed in all cases, we conclude that the number of modes observed with frequencies lower than the lowest frequency phonon vanishes at least as $L^{-1}$. For protocols that generate the $\omega^4$ law (see main text), it vanishes as $L^{-2}$. In any event, we conclude that observing the $\omega^\beta$ tail of the density of vibrational modes in a statistically robust manner in larger systems requires increasingly larger ensembles of glassy samples to be generated, which quickly becomes computationally challanging. 

For the reasons discussed above, we focused on the analysis of systems of $N\!=\!2000, 4000$ and $10,000$ particles. Each ensemble is generated by first producing 10,000 independent equilibrium configurations at the parent temperatures $T_0\!=\!2.00, 0.70$ and $0.60$, and performing an instantaneous quench of each of these independent equilibrium configurations as described in the previous Section. We calculated the first (lowest) 100, 200, and 400 modes for each member of our ensembles of systems of size $N\!=\!2000, 4000$ and $10,000$, respectively, and obtained the distributions $D(\omega)$ of vibrational modes. 

Our results are presented in Fig.~\ref{finite_size_fig}. Our data for systems of size $N\!=\!2000$ and $N\!=\!4000$ do not show any systematic trend, and the respective slopes of $D(\omega)$ at low frequencies do not seem to depend on system size. For the larger systems of $N\!=\!10,000$ the intrusion of the lowest frequency phonon begins to `pull' the distribution upwards, which leads to what could be interpreted as a steeper exponent $\beta$. We reiterate here that we do not aim at accurately determining the numeric value of the exponent $\beta$, but rather identify the trends that it follows as the preparation protocol of our glassy samples is systematically varied. We assert that, within the statistical limitations presented by our data, and within the window of system sizes in which the distributions can be robustly observed, finite size effects do not appear to be significant. 

%This effect appears to be strongest for the $T_0\!=\!2.00$ data, which is consistent with the observation of stronger hybridizations and enhanced delocalization that extend to lower frequencies for the $T_0\!=\!2.00$ ensemble, as evident by the participation ratio data of Fig.~3 in the main text. 

% We do not find significant finite size effects in these data; f

\end{document}